# A search for disordered (glassy) phase in solid $^3$He deformed in situ

A.A. Lisunov, V.A. Maidanov, V.Yu. Rubanskiy, S.P. Rubets, E.Ya. Rudavskii, A.S. Rybalko and V. A. Tikhii

*B.Verkin Institute for Low Temperature Physics and Engineering of the National Academy of Sciences of Ukraine*

*rudavskii@ilt.kharkov.ua*

A disordered (glassy) state has been searched in solid $^3$He deformed in the course of experiment employing precise measurements of pressure. The analysis of the temperature dependence of the crystal pressure measured at a constant volume shows that the main contribution to the pressure is made by the phonon subsystem, the influence of the disordered phase being very weak. Annealing of the deformed crystal does not affect this state. The results obtained differ greatly from the corresponding data for solid $^4$He measured in the region of supersolid effects where a pressure excessive in comparison to the phonon one was registered. The excess pressure had a quadratic dependence on temperature, which is typical of a disordered system. Absence of the excess pressure in solid $^3$He is unclear yet, some speculative interpretations are suggested.

Keywords: solid $^3$He, pressure measurements, supersolid, dislocations.

PACS number(s):67.80.-s, 67.25. du, 67.30. ef

## 1. Introduction

Recently, an unusual behavior of solid $^4$He has been observed in the region where were a prediction of a special state of quantum crystal - supersolidity. The first manifestation of this state was observed in the torsion experiments [1] as a nonclassical rotational moment of inertia. Then an anomalous shear modulus [2], a heat capacity peak [3] and unusual mass transfer [4] were registered in the same temperature region. The fraction of nonclassical rotational inertia was very sensitive to the crystal quality and decreased sharply after annealing [5], which suggests a non-superfluid origin of the supersolid effects [6]. Those effects might be connected with the formation of a disordered (glassy) state in solid helium.

It has been found that the disordered state can be registered suitably using the technique of precise measurements of the crystal pressure [7-9]. The performed experiments showed that in solid $^4$He the pressure in the region of expected supersolidity, apart from the phonon contribution proportional to $T^4$, takes a contribution proportional to $T^2$ which is typical of a disordered (glass) phase. Below T~200mK this contribution was much higher than the phonon one, but the glass phase was destroyed after thorough annealing of the crystal.

Note that these features were observed in solid $^4$He. As concerns the second isotope $^3$He which is a Fermi solid instead of a Bose solid, the first torsion experiments [1] showed absence of the nonclassical rotational inertia. In the experiment [1] solid $^3$He was confined in a porous Vycor glass, therefore the torsion measurements were repeated later in [10] without the porous matrix. The absence of the effect was confirmed which illustrates the importance of quantum statistics. Meantime, a surprising results was obtained in acoustic measurements on solid $^3$He [10]: the shear modulus of $^3$He changed, just like in $^4$He, in the HCP phase but remained normal in the BCC phase. It is likely that the difference in the "supersolid" behavior of the solid helium isotopes is caused not only by the type of statistics, but by the type of the crystal structure, defects or other factors as well.

It was therefore reasonable to search for the disordered state in both phases of $^3$He and compare the results obtained with the corresponding experimental data for $^4$He. The goal of this study was to investigate this problem through precise measurements of pressure of helium crystals at a constant volume. The experimental conditions were similar to those for solid $^4$He [9]. The temperature dependences of the pressure measured after deformation and thorough annealing of the sample were then compared.

## 2. Experimental technique

Like in the case of solid $^4$He, the disordered phase was searched for in solid $^3$He samples deformed in the course of the experiment using the same two-chamber cell detailed in [9]. The samples were grown in one of the chambers in the form of a disk 0.5 mm high and 32 mm in diameter. The samples cooled below 100 mK were deformed using other (control) chamber. For this purpose, liquid $^4$He was condensed into the control chamber, which permitted cyclic variations of the pressure in it from 0 to ~ 25 atm. The change in the pressure of the solid $^3$He sample was ~ 6 atm. Generally, each sample was put through five deformation cycles. The kinetics of pressure variations in the sample and corresponding temperature response are shown in Fig. 1.

The first spike of the temperature is due to the condensation heat released while filling helium to the control chamber. The subsequent slight rises of temperature on each change (decrease or increase) in the pressure are caused by the viscous heating in the filling line.

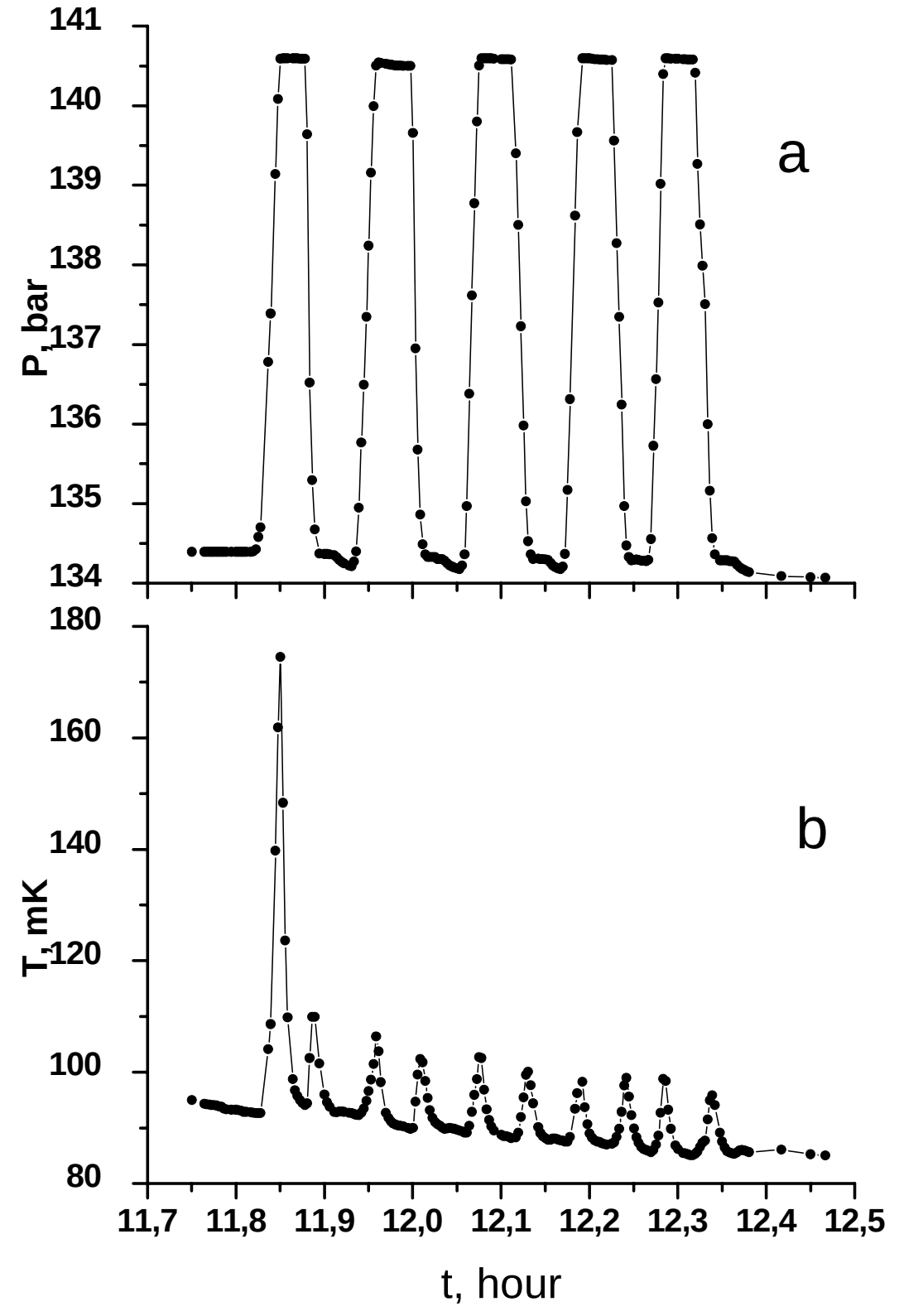

Fig. 1.Kinetics of pressure (a) an temperature (b) variations in the course of sample deformation.

After the deformation process was completed, the pressure in the control chamber was released to zero and $^4$He was pumped out with an adsorption pump for 4-6 hours. Note that the pressure of the deformed sample practically came back to its starting value through the process of relaxation was rather slow (up to 24-48 hours).

The starting $^3$He contained ~ 0.25% of the $^4$He impurity and was rectified in a void column (its basic structure is described in [11]). Normally, such columns yield 0.999997 pure $^3$He. For lack of an accurate analysis, the upper limit of possible $^4$He impurities was estimated indirectly. The crystal grown from the purified $^3$He on cooling to ~ 65 mK had no evidence of phase separation, which suggests the $^4$He concentration in the sample was below 10ppm.

The samples were grown by the capillary-blockage technique and investigated in an interval of 100-600 mK. The pressure of the sample was measured with Straty-Adams capacitive gauge and a precise GR-1615A capacitance bridge, the accuracy being ±3Pa. Five samples were investigated in the region of molar volumes ~ 19-22 cm$^3$/mole, which corresponds to both BCC and HCP phases of $^3$He. The measurements were performed on deformed samples before and after their thorough annealing for 20-24 hours near the melting temperature.

## 3. Analysis of pressure vs. temperature dependence

The temperature dependence of pressure P(T) was measured on stepwise decreasing and increasing the temperature. The measurements were repeated several times on cooling and heating. The primary experimental results are illustrated in Fig. 2.

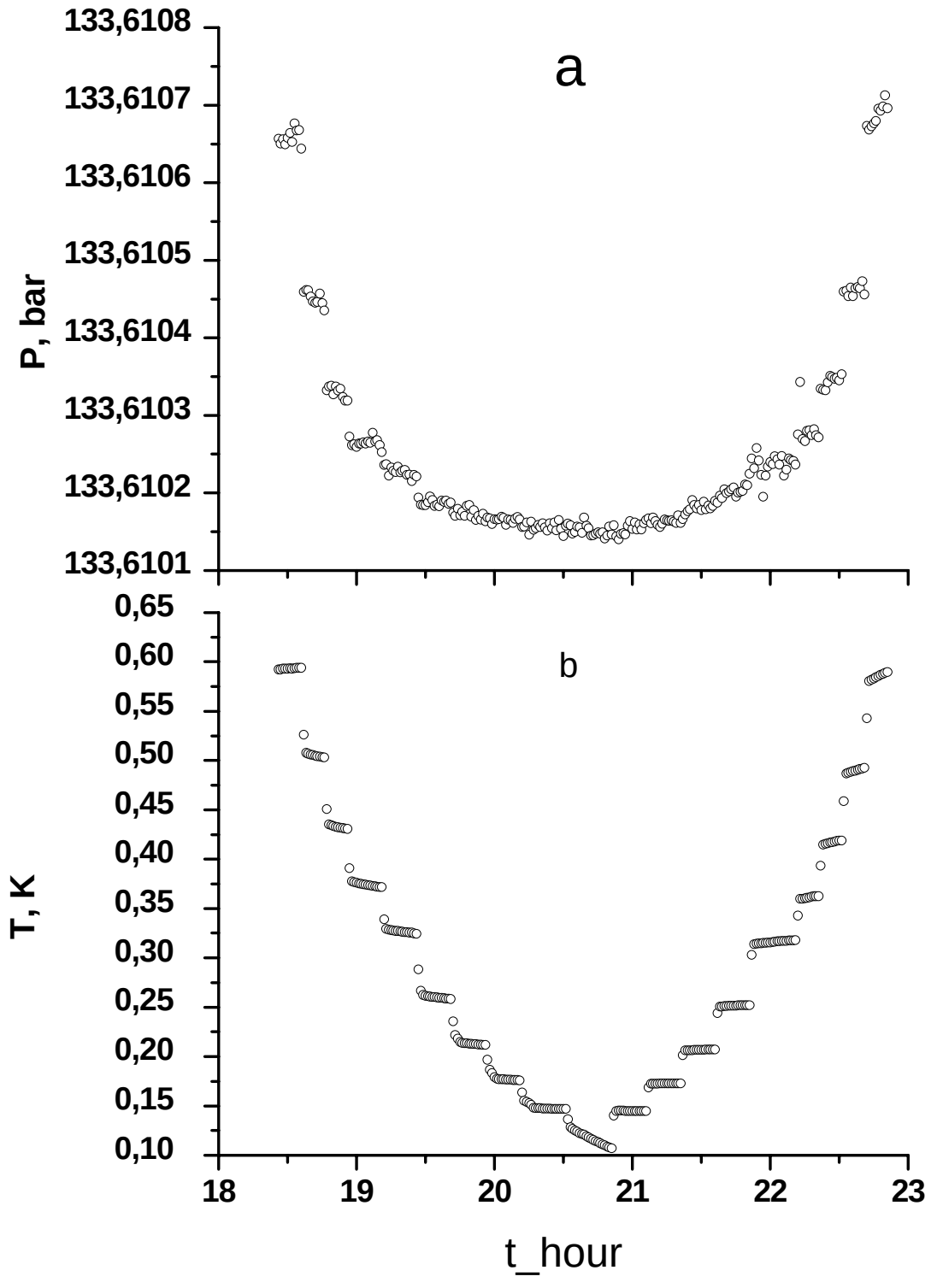

Fig. 2. Typical pressure variations during stepwise cooling and heating

Since vacancy excitations have a negligible effect on pressure in this temperature region, the dependences P(T) obtained on different samples were approximated by the relation

$$P(T)=P_0+P_{ph}(T)+P_d(T), \qquad (1)$$

where $P_0$ is the crystal pressure at T=0, $P_{ph}(T)=a_{ph}T^4$ is the phonon contribution to the pressure. The last term describes the contribution of a disordered phase that might be generated by deformation of the $^3$He crystal. Normally, the temperature dependence of pressure in this phase is $P_d(T)=a_dT^2$. It is then convenient to re-write Eq. (1) as

$$[P(T) - P_0] / T^2 = a_d + a_{ph}T^2 \qquad (2)$$

as a result, the dependences P(T) replotted in the coordinates $[P(T) - P_0] / T^2$ vs. $T^2$ permit a straightforward estimation of the fitting parameters $a_d$ and $a_{ph}$.

Such a dependence of one of the $^3$He samples is shown as an example in Fig. 3. As was expected, these dependences are straight lines. Their sloped determine the parameter $a_{ph}$, and the intercept on the ordinate axis yields the parameter $a_d$.

The curves in Fig. 3 were taken on a deformed samples before and after annealing. It is seen that both the curves (1 and 2) coincide within the experimental data scatter and give $a_d$ close to zero, which suggests that a disordered phase is practically absent in the crystal.

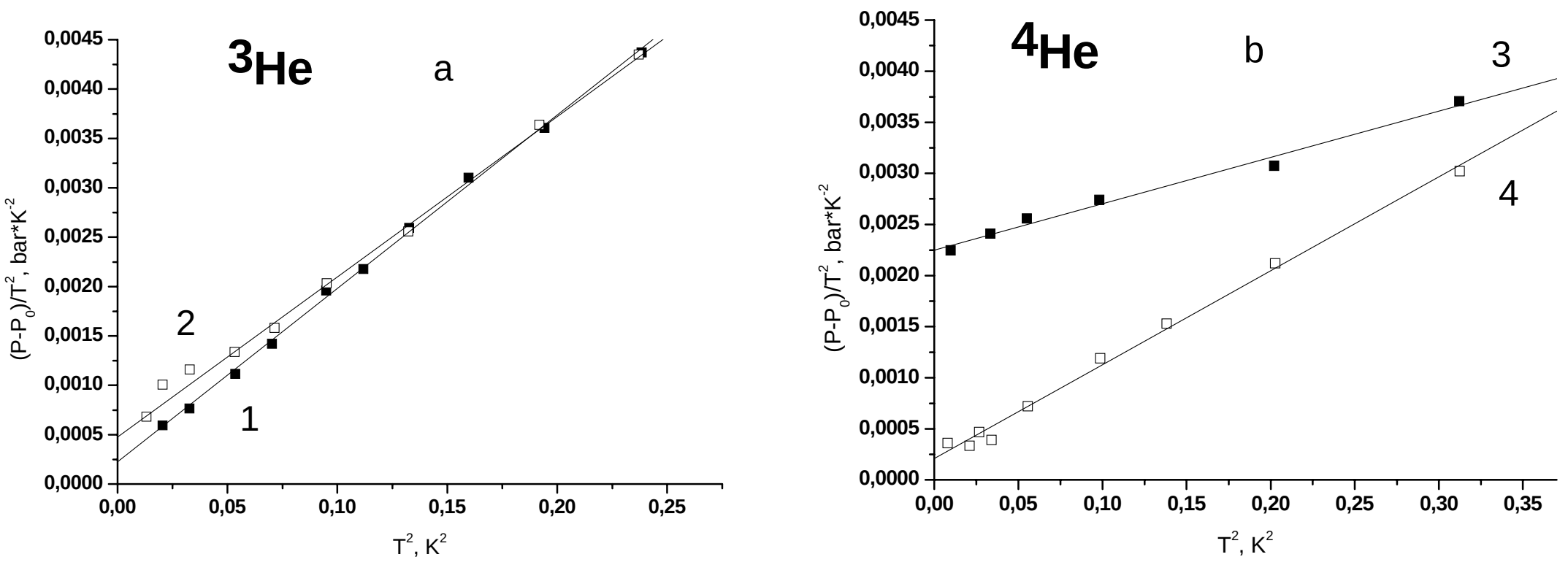

Fig. 3. Dependences $[P(T) - P_0] / T^2$ vs. $T^2$ of deformed solid He samples:

a – solid $^3$He before (1) and after (2) annealing;

b – solid $^4$He before (3) and after (4) annealing

For comparison, Fig. 3 carries the corresponding dependences for a solid $^4$He sample, which show distinctly the presence of a disordered (glassy) phase in the deformed crystal. It is described by the parameter $a_d$=0.022 bar/K$^2$ (curve 3). After thorough annealing (curve 4) $a_d$ is close to zero, i.e. the contribution of the disordered phase is very small. Thus, unlike the case of $^4$He, deformation of a solid $^3$He crystal does not cause the formation of appreciable disordered phase.

According to the experimental findings, this is true for all of the five $^3$He samples (Fig. 4).

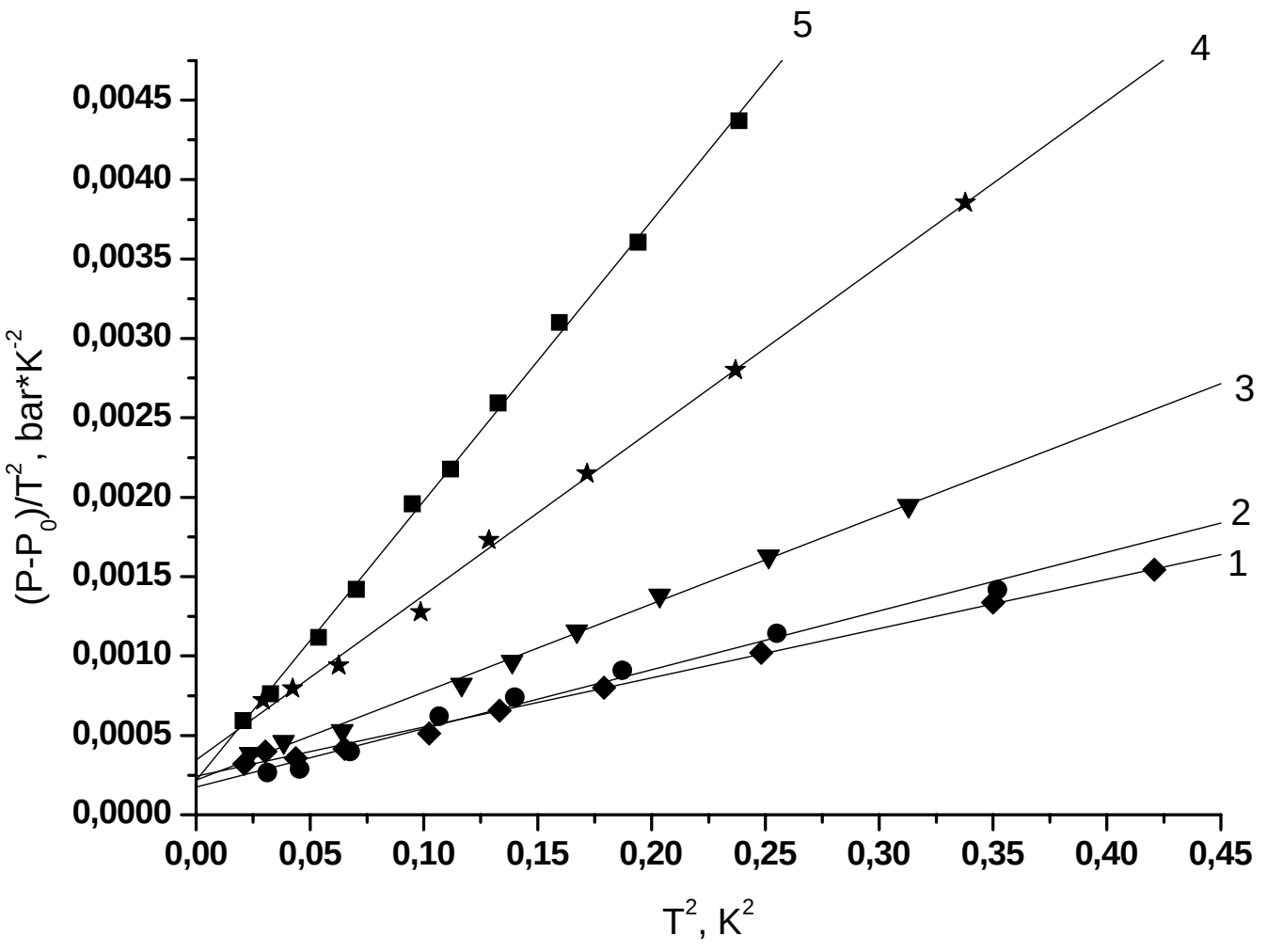

Fig. 4. Dependences *[P(T) –P₀] / T²* vs. $T^2$ for deformed solid $^3$He samples of various molar volumes (cm$^3$/mole): 1-19.2, 2-19.0, 3-19.8, 4-20.7, 5-22.2

It should be emphasized that curves 1 and 2 refer to the HCP phase of $^3$He, while curves 3-5 describes the BCC phase, i.e. the result is independent of the crystalline structure of the solid $^3$He.

## 4. Conclusions

The reason why deformation of a solid $^3$He crystal unlike a $^4$He one, does not cause a substantial quantity of the disordered phase is so far physically obscure and allows only speculative interpretations.

Normally, deformation gives rise to a system of dislocations in a crystal, which may therefore be considered as a key factor in the above effects. The behavior of the pressure in deformed $^4$He [9] was analyzed through calculating the contribution of dislocations to the thermodynamics properties of the crystal [12], which enabled the authors to describe the experimental results and to estimate the density of dislocations in the crystal. In the case of $^3$He the dislocations in the crystal become more mobile because of the larger amplitudes of zero-point oscillations which may be obstacle to forming and maintaining the required density of dislocations.

The experimental results for $^4$He [9] are also in qualitative agreement with the glass model of two-level tunnel states [13-15], which predicts a quadratic temperature dependence of pressure. The analysis of the experiment within this model allows estimation of the density of two-level states. Since the microscopic atom model of such two-level states has not been proposed yet the applicability of this approach for describing the solid $^3$He behavior remains open.

In connection with different types of quantum statistic of $^3$He and $^4$He, the scenario of quasi-one-dimensional superfluidity along the dislocation lines [16,17] in a Bose system should not be ruled out either. The contribution of such system to pressure is also proportional to $T^2$ [18]. It is natural that this scenario is impossible in solid $^3$He.

The conclusive interpretation of the results obtained in this study calls for additional theoretical and experimental investigations.

The study was partially supported by STCU (Project 5211). The authors are grateful to S.I. Shevchenko, V.D. Natsik, S.N. Smirnov for helpful discussions.